\documentclass[prl,twocolumn,amsmath,amssymb,superscriptaddress,preprintnumbers]{revtex4}
\usepackage[pdftex]{graphicx}
\usepackage[outdir=./]{epstopdf}
\usepackage{enumerate,ulem}
\newcommand \beq{\begin{eqnarray}}
\newcommand \eeq{\end{eqnarray}}

\newcommand{\bm}[1]{\boldsymbol{#1}}

\newcommand{\bfr}{\bm{r}}

\newcommand{\bfM}{\bm{M}}
\newcommand{\bfp}{\bm{p}}
\newcommand{\bfsigma}{\bm{\sigma}}

\newcommand{\sech}{\mathrm{sech}\,}

\begin{document}

\title{Charge pumping induced by magnetic texture dynamics in Weyl semimetals}
\author{Yasufumi Araki}
\affiliation{Institute for Materials Research, Tohoku University, Sendai 980--8577, Japan}
\affiliation{Frontier Research Institute for Interdisciplinary Sciences, Tohoku University, Sendai 980--8578, Japan}
\author{Kentaro Nomura}
\affiliation{Institute for Materials Research, Tohoku University, Sendai 980--8577, Japan}

\begin{abstract}
Spin--momentum locking in Weyl semimetals correlates the orbital motion of electrons with background magnetic textures. We show here that the dynamics of a magnetic texture in a magnetic Weyl semimetal induces a pumped electric current that is free from Joule heating. This pumped current can be regarded as a Hall current induced by axial electromagnetic fields equivalent to the magnetic texture. Taking a magnetic domain wall as a test case, we demonstrate that a moving domain wall generates a pumping current corresponding to the localized charge.
\end{abstract}


\maketitle

\textit{Introduction}---Magnetic textures, such as domain walls (DWs), skyrmions, spin spirals, etc., are currently attracting a great deal of interest in condensed matter physics. In the context of spintronics, these magnetic textures are expected to assume an integral role as information carriers in next-generation devices and in switching devices driven by electric and spin currents \cite{Zutic_Fabian_DasSarma,Brataas_Bauer_Kelly,Tatara_Kohno_Shibata}.
In particular, the dynamical properties of such magnetic textures,
which are coupled to the spins of conduction electrons, are the focus of intense efforts to control and detect them efficiently, and promise a wide range of future applications \cite{Spin_current}. Depending on the particular context, different dynamical perspectives can be used to describe the coupling between magnetic textures and conduction electrons. One may view the spin-transfer torque arising from the spins of conduction electrons as primarily responsible for driving the dynamics of magnetic textures \cite{Ralph}, or conversely, see the dynamics of a magnetic texture as inducing an external force on conduction electrons through changes in the electron Berry phase, which is known as the spin motive force \cite{Barnes}.

In this work, we propose that the dynamics of magnetic textures in Weyl semimetals (WSMs) can invoke electric charge pumping free from backscattering in a manner that is distinct from that induced by the spin motive force.
WSMs form a class of topological materials characterized by a conical band structure and pair(s) of band-touching points (Weyl points) isolated from each other in the bulk Brillouin zone \cite{Young_2012,Wan_2011,Burkov_2011,Zyuzin_2012}.
{This} ``Weyl cone'' structure arises from band inversion due to strong spin--orbit coupling (SOC), and is associated with significant electron spin--momentum locking around the nodal points. For such spin--momentum-locked electrons, the exchange coupling to the background magnetic texture is analogous to a fictitious vector potential, which is referred to as an ``axial vector potential'' \cite{Liu_2013}.
In the context of this analogy, we can then observe a ``Hall current'' free from Joule heating that is induced by the axial magnetic and electric fields corresponding to the dynamics of the background magnetic texture. This ``Hall effect'' accounts for the charge pumping mechanism proposed here.

{The significance of magnetic textures in WSMs has been discussed in several recent studies, and based on the features of the electron spin--momentum locking, it was proposed that the correlation between the magnetic moments (mediated by the Weyl electrons) exhibits longitudinal anisotropy, distinguishing it from that due to ordinary isotropic Ruderman--Kittel--Kasuya--Yosida (RKKY) interactions}
\cite{Hosseini_2015,Chang_2015,Araki_corr}.
{Such anisotropic correlations then give rise to the formation of nontrivial topological magnetic textures in WSMs. Moreover, once a magnetic DW is formed in a WSM, it is accompanied by a certain amount of electric charge and an equilibrium current localized to the DW}, no matter how the DW was introduced into the WSM \cite{Araki_DW,Grushin}.
{While the theories proposed in these studies principally account for the characteristics of static magnetic textures in WSMs, the dynamical properties of magnetic textures in WSMs are much less understood and a better understanding of such properties is required for the ability to read and write information in future devices. The charge pumping effect discussed in this work is one such dynamical property attributed to magnetic textures in WSMs. Moreover, since it arises as a Hall current due to axial electromagnetic fields, it is free from Joule heating and could lead to reduced power consumption in future spintronic applications.}

\textit{Axial vector potentials}---The formation of the WSM phase requires the breaking of either time-reversal symmetry (TRS) or spatial inversion symmetry so that the degeneracy of the Weyl cones is lifted. TRS-breaking in WSMs is typically realized by the introduction of magnetic order in the system, such as the hypothetical case of ferromagnetic order in Co-based Heusler alloys \cite{Bernevig_2016,Hasan_2016}, or chiral antiferromagnetic order in Mn-based materials (e.g. $\mathrm{Mn}_3 \mathrm{Sn}$) \cite{Kubler_2016,Yang_2017,Ito}. The breaking of TRS shifts the positions of the Weyl nodes in momentum space, and this effect can be regarded as an “emergent vector potential'' for each Weyl node \cite{Liu_2013}. As far as the low-energy phenomena are concerned, one can then rely on this idea of an effective vector potential to treat the temporal and spatial variations of the magnetization in the system more efficiently.

Here, we consider a minimal model of a WSM exhibiting ferromagnetic order,
with a pair of Weyl cones dispersed isotropically around each Weyl node.
The low-energy electrons (around the Weyl nodes) can then be described by the continuum Hamiltonian
\begin{align}
\mathcal{H} = s v_\mathrm{F} (\bfp \cdot \bfsigma) - J \bfM(\bfr,t) \cdot \bfsigma,
\end{align}
where $s = \pm 1$ denotes the chirality of each Weyl node,
$\bfsigma = (\sigma_x,\sigma_y,\sigma_z)$ are the Pauli matrices corresponding to the electron spin degrees of freedom, $v_\mathrm{F}$ is the Fermi velocity, and $\bfp = -i\boldsymbol{\nabla}$ is the electron momentum operator. For convenience, we have taken $\hbar=1$ here. The second term in this Hamiltonian describes the exchange coupling between the electron spin $\bfsigma$ and the local magnetic texture $\bfM(\bfr,t)$, with coupling constant $J$.
Then, provided that our Weyl-cone approximation is valid, the local magnetic texture $\bfM(\bfr,t)$ can be viewed as a U(1) axial gauge potential $\boldsymbol{A}_5(\bfr,t) = (J/v_\mathrm{F} e)\bfM(\bfr,t)$ coupled to the electrons,
and in terms of which our Hamiltonian can be written as
\begin{align}
\mathcal{H} = s v_\mathrm{F} \left[ \bfp - s e \boldsymbol{A}_5(\bfr,t) \right]\cdot\bfsigma.
\end{align}
In contrast to a normal vector potential $\boldsymbol{A}$, the axial vector potential $\boldsymbol{A}_5$ couples to different chirality modes $(s=\pm)$ with opposite sign and is not subject to Maxwell's equations.
As we shall see in the following sections (and by analogy to normal vector potentials), it is this axial vector potential that is responsible for the proposed electron transport.

\begin{table}[tbp]
\caption{Classification of currents induced by normal and axial electromagnetic fields (EMFs).
The current induced by axial EMFs is evaluated with $\mu_5=0$.}
\begin{tabular}{|c||c|c|}
\hline
 & Normal EMFs: $(\boldsymbol{E}, \boldsymbol{B})$ & Axial EMFs: $(\boldsymbol{E}_5, \boldsymbol{B}_5)$ \\
\hline
\hline
Drift & $\boldsymbol{j}^\mathrm{(D)} = \sigma_\mathrm{D} \boldsymbol{E}$ & $\boldsymbol{j}^\mathrm{(D)} = 0$ \\
AHE & $\boldsymbol{j}^\mathrm{(A)} = \sigma_\mathrm{A} \hat{\boldsymbol{M}} \times \boldsymbol{E}$ & $\boldsymbol{j}^\mathrm{(A)} = 0$ \\
CME & $\boldsymbol{j}^\mathrm{(C)} = (e^2/2\pi^2) \mu_5 \boldsymbol{B}$ & $\boldsymbol{j}^\mathrm{(C)} = (e^2/2\pi^2) \mu \boldsymbol{B}_5$ \\
RHE & $\boldsymbol{j}^\mathrm{(H)} = \sigma_\mathrm{H} \hat{\boldsymbol{B}} \times \boldsymbol{E}$ & $\boldsymbol{j}^\mathrm{(H)} = \sigma_\mathrm{H} \hat{\boldsymbol{B}}_5 \times \boldsymbol{E}_5$ \\
\hline
\end{tabular}
\label{table:current}
\end{table}

\textit{Field-induced current}---Before considering the proposed charge pumping mechanism, let us first review the different kinds of electric current induced by real EMFs.
Currents induced by EMFs in a WSM can be classified based on their linear response to an electric field $\boldsymbol{E}$ and/or magnetic field $\boldsymbol{B}$ (see Table \ref{table:current}).
If only an $\boldsymbol{E}$-field is applied to the WSM,
a longitudinal drift current, $\boldsymbol{j}^\mathrm{(D)} = \sigma_\mathrm{D} \boldsymbol{E}$, is induced, where $\sigma_\mathrm{D}$ is the longitudinal conductivity of a pair of Weyl cones. If TRS is broken in the WSM by the presence of magnetization $\boldsymbol{M}$, an anomalous Hall effect (AHE) is also present \cite{Zyuzin_2012_2,Goswami_2013,Burkov_2014,Burkov_2014_2}, 
and drives the transverse current $\boldsymbol{j}^\mathrm{(A)} = \sigma_\mathrm{A} \hat{\boldsymbol{M}} \times \boldsymbol{E}$,
where the anomalous Hall conductivity is given by $\sigma_\mathrm{A} = (e/2\pi^2)(J|\boldsymbol{M}|/v_\mathrm{F})$ \cite{hat}.

On the other hand, if only a magnetic field $\boldsymbol{B}$ is applied to the system, it induces Landau quantization with a cyclotron frequency $\omega_c = v_\mathrm{F} \sqrt{2e B}$.
Nonzero Landau levels (LLs) then appear symmetrically about the zero energy due to particle--hole symmetry, and the zeroth LL is linearly dispersed along the magnetic field. As the dispersion direction of the zeroth LL for each Weyl node depends on the chirality $s$ \cite{K-Y_Yang}, it only contributes to the net current $\boldsymbol{j}^\mathrm{(C)} = (e^2/2\pi^2) \mu_5 \boldsymbol{B}$
if there is a chemical potential imbalance $\mu_5$ between the two Weyl nodes. This effect is known as the chiral magnetic effect (CME) and accounts for the negative magnetoresistances observed in WSMs \cite{Zyuzin_2012_2,Vilenkin,Fukushima_2008,Kharzeev_2008,Kharzeev_2014}.

Finally, a combination of $\boldsymbol{E}$ and $\boldsymbol{B}$ induces a Hall current perpendicular to both, $\boldsymbol{j}^\mathrm{(H)} = \sigma_\mathrm{H} \hat{\boldsymbol{B}} \times \boldsymbol{E}$,
which we have called the regular Hall effect (RHE) to distinguish it from the AHE. The regular Hall conductivity $\sigma_\mathrm{H}$ depends on both the field strength and the amount of disorder present in the system.
If the level broadening arising from disorder obscures the LL spacing,
i.e., if the cyclotron frequency $\omega_c$ is smaller than the relaxation rate $1/\tau$, the transport coefficients can be estimated in the classical limit using semiclassical (Boltzmann) transport theory \cite{Nandy}.
The zero temperature Hall conductivity is then given by$\sigma_\mathrm{H(c)} = -({\tau^2 e^3 \mu}/{3\pi^2}) |\boldsymbol{B}|$ at the lowest order in $\boldsymbol{B}$, where $\mu$ is the electron chemical potential measured from the Weyl nodes. On the other hand, in the quantum limit where the disorder is dilute and the LLs can be regarded as well separated $(\omega_c \tau \gg 1)$, the Hall current can be effectively described by the ``quantum Hall effect''. If the Fermi level $\mu$ lies just slightly beyond the charge neutrality point so that it does not cross the higher LLs, only the zeroth LL contributes to the Hall current. The Hall conductivity then reduces to the universal value
\begin{align}
\sigma_\mathrm{H(q)} = \frac{e^2}{2\pi^2} \frac{\mu}{v_\mathrm{F}},
\end{align}
which can be derived from the quantum Hall conductivity in 2D Dirac systems such as graphene.

\textit{Charge pumping induced by magnetic texture dynamics}---As we have outlined above, in order to consider the effect of magnetic texture dynamics on the electron transport, we can rely on the idea of axial EMFs. Specifically, the dynamics of the magnetic texture, i.e., $\bfr$- and $t$-dependences in the axial vector potential $\boldsymbol{A}_5$, are equivalent to axial electric and magnetic fields, $\boldsymbol{E}_5$ and $\boldsymbol{B}_5$, given by
\begin{align}
\boldsymbol{E}_5(\bfr,t) &= -\partial_t \boldsymbol{A}_5(\bfr,t) = -\frac{J}{v_\mathrm{F} e} \partial_t \boldsymbol{M}(\bfr,t) \\
\boldsymbol{B}_5(\bfr,t) &= \boldsymbol{\nabla} \times \boldsymbol{A}_5(\bfr,t) = \frac{J}{v_\mathrm{F} e} \boldsymbol{\nabla} \times \boldsymbol{M}(\bfr,t),
\end{align}
respectively. The electron transport induced by the magnetic texture dynamics
can then be treated in terms of these axial EMFs, thus enabling its evaluation in similar fashion to normal EMFs, making the overall discussion quite simple. {As we shall see in the following, the axial electric field $\boldsymbol{E}_5$ drives an ``axial current'' comprising a pair of currents flowing oppositely to each other at the two Weyl nodes and thus yielding no net current, while a net current is induced if it is accompanied by an axial magnetic field $\boldsymbol{B}_5$.} Here, we note that we have neglected intervalley scattering processes, so that the electron transport for each Weyl node could be treated separately.

{As long as the magnetic texture dynamics are sufficiently slow and ``adiabatic''}, the axial electric field $\boldsymbol{E}_5$ is so weak
that its nonlinear effect can be safely discarded. With this assumption, it then simply induces a drift current and an anomalous Hall current for each Weyl node flowing in opposite directions to one another, i.e., the axial current \cite{Taguchi}. As such, it contributes no net current unless there is an imbalance in the carrier densities $(\mu_5 \neq 0)$.
On the other hand, the RHE contribution is same as that induced by normal EMFs, i.e., $\boldsymbol{j}^\mathrm{(H)} = \sigma_\mathrm{H} \hat{\boldsymbol{B}}_5 \times \boldsymbol{E}_5$, since both $\boldsymbol{E}_5$ and $\boldsymbol{B}_5$ couple to each chiral mode with opposite signs, driving the Hall current for each Weyl node in the same direction. As long as the disorder is weak enough compared with the level spacing, the induced Hall current can be estimated in the quantum limit
in similar fashion to the case for real $\boldsymbol{B}$ and $\boldsymbol{E}$, yielding
\begin{align}
\boldsymbol{j}^\mathrm{(H)} = \frac{e^2}{2\pi^2}\frac{\mu}{v_\mathrm{F}}\hat{\boldsymbol{B}}_5 \times \boldsymbol{E}_5, \label{eq:Hall-current}
\end{align}
which is independent of the field strength $|\boldsymbol{B}_5|$.
Moreover, since the zeroth LLs of the two Weyl nodes are dispersed in the same direction along $\boldsymbol{B}_5$ \cite{Araki_DW,Grushin,Liu_2013,Pikulin}, a finite chemical potential leads to the net current
\begin{align}
\boldsymbol{j}^\mathrm{(C)} = \frac{e^2}{2\pi^2}\mu \boldsymbol{B}_5, \label{eq:CME-current}
\end{align}
which we identify as the chiral \textit{axial} magnetic effect (CAME),
i.e., the axial counterpart of the CME.
Therefore, the total current $\boldsymbol{j}_\mathrm{ind}$ induced by $\boldsymbol{E}_5$ and $\boldsymbol{B}_5$ in WSMs is given (up to the linear response in $\boldsymbol{E}_5$) by the sum of $\boldsymbol{j}^\mathrm{(H)}$ and $\boldsymbol{j}^\mathrm{(C)}$.

We should note that typical axial EMFs arising from magnetic textures
are spatially inhomogeneous. However, if the magnetic texture is sufficiently smooth over the relevant length scales (e.g., the electron's mean free path), the axial EMFs can be regarded as ``locally'' uniform and we can consider the properties of the electron transport in the ballistic limit. In such cases, we can use Eqs.~(\ref{eq:Hall-current}) and (\ref{eq:CME-current}) to estimate the local current distribution. In the absence of normal EMFs $\boldsymbol{E}$ and $\boldsymbol{B}$, the axial anomaly between the chiral modes (see \cite{Adler,Bell_Jackiw,Nielsen_Ninomiya}) does not violate the conservation of charge \cite{Liu_2013}, and we can use the charge conservation relation
\begin{align}
\partial_t \rho_\mathrm{pump}(\bfr,t) = -\boldsymbol{\nabla}\cdot \boldsymbol{j}_\mathrm{ind}(\bfr,t), \label{eq:charge-conservation}
\end{align}
to estimate the electric charge $\rho_\mathrm{pump}(\bfr,t)$ pumped
by the magnetic texture dynamics $\boldsymbol{M}(\bfr,t)$
via the axial field-induced current $\boldsymbol{j}_\mathrm{ind}$.
Since the CAME part, $\boldsymbol{j}^\mathrm{(C)} \propto \mu (\boldsymbol{\nabla} \times \boldsymbol{M})$, is divergence-free whenever the chemical potential is uniform, only the regular Hall current given by Eq.~(\ref{eq:Hall-current}) is responsible for the ensuing charge dynamics:
\begin{align}
\partial_t \rho_\mathrm{pump} = \frac{e^2}{2\pi^2}\frac{\mu}{v_\mathrm{F}} \left[ \hat{\boldsymbol{B}}_5 \cdot (\boldsymbol{\nabla}\times\boldsymbol{E}_5) - \boldsymbol{E}_5 \cdot (\boldsymbol{\nabla}\times\hat{\boldsymbol{B}}_5)\right] . \label{eq:pumping}
\end{align}
This equation is the key result of this work, and directly relates the magnetic texture dynamics (via $\boldsymbol{E}_5$ and $\boldsymbol{B}_5$)
to the pumped charge $\rho_\mathrm{pump}$.

We note that if the spatial variation of $\boldsymbol{M}$ is coplanar, the axial magnetic field $\boldsymbol{B}_5$ is uniform, i.e., $\hat{\boldsymbol{B}}_5$ is homogeneous over the whole system.
In this case, the second term in Eq.~(\ref{eq:pumping}) vanishes and we obtain
$\partial_t \rho_\mathrm{pump} = -({e^2}/{2\pi^2})({\mu}/{v_\mathrm{F}})(\hat{\boldsymbol{B}}_5 \cdot \partial_t \boldsymbol{B}_5) = -({e^2}/{2\pi^2})({\mu}/{v_\mathrm{F}}) \partial_t |\boldsymbol{B}_5|$,
where we have used the relation $\boldsymbol{\nabla}\times\boldsymbol{E}_5 = \boldsymbol{\nabla}\times(-\partial_t\boldsymbol{A}_5) = -\partial_t \boldsymbol{B}_5$.
Thus, we obtain a further simplified relation for this restricted case
\begin{align}
\rho_\mathrm{pump}(\bfr,t) = -\frac{e^2}{2\pi^2}\frac{\mu}{v_\mathrm{F}} |\boldsymbol{B}_5(\bfr,t)| + \mathrm{const.} \label{eq:induced-charge}
\end{align}
which implies that an axial magnetic flux (i.e., the curl of the magnetization) induces localized electric charge in a WSM, irrespective of its orientation.

\textit{Example: Magnetic domain wall}---In order to establish the validity of the relations presented above, let us consider a moving magnetic DW in a WSM as a typical example. We construct a DW of width $2w$ in the $yz$-plane, separating two regions of an infinite system with magnetizations $\bfM(x \rightarrow \pm \infty) = \pm M_0 \boldsymbol{e}_z$, and then set the DW in motion with velocity $V_\mathrm{DW}$ in the $x$-direction by hand. The resulting magnetic texture is then given by
\begin{align}
\bfM(\bfr,t) = \bfM(x-V_\mathrm{DW}t) = M_0 
\begin{pmatrix}
\lambda_x \sech\xi(x,t) \\ \lambda_y \sech\xi(x,t) \\ \tanh\xi(x,t)
\end{pmatrix},
\end{align}
where $\xi(x,t) \equiv (x-V_\mathrm{DW} t)/w$ denotes the relative position from the center of the DW, rescaled by the DW width. The set of parameters $(\lambda_x,\lambda_y)$ characterizes the texture of the DW, where a DW with a coplanar magnetic texture within the $xz$-plane (i.e., a N\'{e}el DW) corresponds to $(\lambda_x,\lambda_y) = (\pm 1,0)$,
while a DW with a helical magnetic texture twisting in the $yz$-plane
(i.e., a Bloch DW) is given by $(\lambda_x,\lambda_y) = (0,\pm 1)$.
In reality, the particular type of DW assumed in such a system is microscopically chosen by the mechanism stabilizing the DW (e.g., the Dzyaloshinskii--Moriya interaction), but we do not go into such details here and simply consider an arbitrary type of DW.

\begin{figure}[tbp]
 \includegraphics[width=7cm]{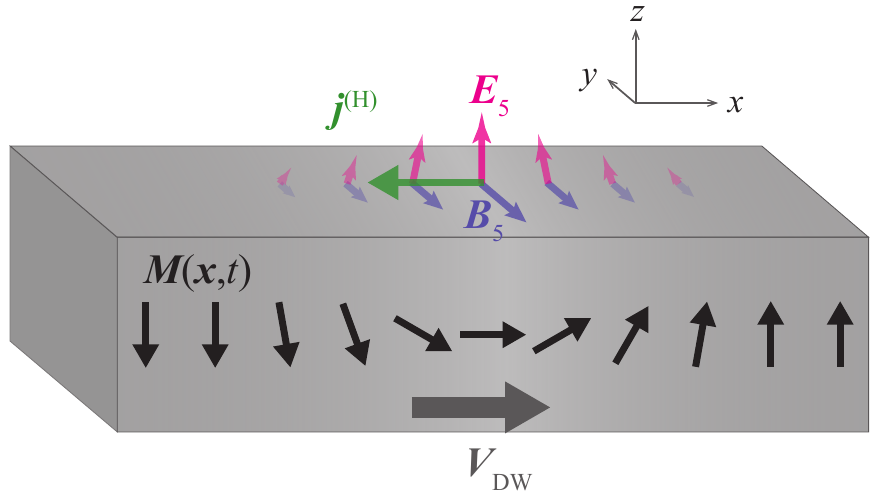}
\caption{Schematic picture showing the axial EMFs $(\boldsymbol{E}_5,\boldsymbol{B}_5)$ and the induced Hall current $\boldsymbol{j}^\mathrm{(H)}$, along with a N\'e{e}l domain wall moving with velocity $V_\mathrm{DW}$.}
\label{fig:configuration}
\end{figure}

In order to estimate the current $\boldsymbol{j}_\mathrm{ind}$ induced by the DW's motion, we consider the axial gauge field $\boldsymbol{A}_5 = (J/v_\mathrm{F} e)\boldsymbol{M}$. The axial EMFs are then given by
\begin{align}
\boldsymbol{E}_5 = \frac{JV_\mathrm{DW}}{e v_\mathrm{F} w} \boldsymbol{M}'(\xi), \quad
\boldsymbol{B}_5 = \frac{J}{e v_\mathrm{F} w} \boldsymbol{e}_x \times \boldsymbol{M}'(\xi),
\end{align}
where $\boldsymbol{M}'(\xi) \equiv d \boldsymbol{M}(\xi)/d\xi$.
The magnitude and orientation of the axial EMFs are shown schematically in Fig.~\ref{fig:configuration}. In the case of a DW with $w = 100~\mathrm{nm}$ and $V_\mathrm{DW} = 100~\mathrm{m/s}$ in a magnetic WSM with $v_\mathrm{F} = 10^6~\mathrm{m/s}$ and $JM_0 = 100~\mathrm{meV}$, the strength of the axial fields at the center of the DW are given by $|\boldsymbol{E}_5| = 1~\mathrm{V/cm}$ and $|\boldsymbol{B}_5| = 1~\mathrm{T}$.

Then, using the axial EMFs presented above and assuming that the chemical potential $\mu$ is slightly above the point of charge neutrality so that the quantum limit is valid, the regular Hall current corresponding to the charge pumping effect can be estimated from Eq.~(\ref{eq:Hall-current}) to find
\begin{align}
\boldsymbol{j}^\mathrm{(H)} = -\frac{e}{2\pi^2}\frac{JV_\mathrm{DW} \mu}{v_\mathrm{F}^2 w} \left[ |\boldsymbol{M}'_\perp|\boldsymbol{e}_x + M'_x (\boldsymbol{e}_x \times \hat{\boldsymbol{B}}_5) \right],
\end{align}
where $\boldsymbol{M}_\perp = (0,M_y,M_z)$.
The first term in the above equation represents the longitudinal current flowing along the moving DW, while the second term is the transverse current flowing parallel to the DW, which is independent of $(y,z)$ and does not affect the charge conservation relation [Eq.~(\ref{eq:charge-conservation})] since it is divergence-free.
From this, we see that the charge pumping predominantly occurs close to the DW center, where the most drastic variation in $\boldsymbol{M}_\perp(x,t)$ occurs.

The amount of electric charge pumped along with the DW can be derived from the induced current using the charge conservation relation. Since both of the DW's $x$- and $t$-dependences are characterized by a single variable $\xi = (x-V_\mathrm{DW} t)/w$, the differential operators on both sides of Eq.~(\ref{eq:charge-conservation}) are easy to treat, and lead to the charge distribution
\begin{align}
\rho_\mathrm{pump}(x,t) = \frac{1}{V_\mathrm{DW}}j_x^\mathrm{(H)}(x,t) = -\frac{e}{2\pi^2}\frac{J \mu}{v_\mathrm{F}^2 w}|\boldsymbol{M}'_\perp|.
\end{align}
The net amounts of charge per unit area pumped by N\'{e}el and Bloch DWs are then given by
\begin{align}
q_\mathrm{pump}^\mathrm{(Neel)} = -\frac{e}{\pi^2} \frac{JM_0}{v_\mathrm{F}^2}\mu , \quad
q_\mathrm{pump}^\mathrm{(Bloch)} &= -\frac{e}{2\pi} \frac{JM_0}{v_\mathrm{F}^2}\mu,
\end{align}
respectively. These net charges are independent of the DW width $w$, which implies that the charge pumping is indeed a topological effect. Moreover, in the case of the N\'{e}el wall, $q_\mathrm{pump}^\mathrm{(Neel)}$ successfully accounts for the same amount of localized charge obtained by exactly counting the number of bound states that was presented in previous work \cite{Araki_DW}, which provides a guarantee of the validity of the quantum limit employed in this work. Furthermore, the charge pumping picture presented here can be used for any other type of DW, as long as the DW texture is sufficiently sharp so that the quantum limit approximation can be applied. As the pumping current discussed here can be effectively described as a quantum Hall effect, it is free from energy loss by Joule heating and thus distinct from the drift current arising from the spin motive force.

\textit{Conclusions and Outlook}---We have discussed the relation between the dynamics of magnetic textures and charge pumping in magnetic WSMs. Since the coupling between the magnetization and Weyl electrons may be viewed in terms of an axial gauge potential, the curl and time derivative of the magnetic texture correspond to axial magnetic and electric fields, respectively. The main message of this work [Eqs.~(\ref{eq:Hall-current}) and (\ref{eq:pumping})] is that these axial EMFs give rise to a regular Hall current, which can be regarded as a pumping current induced by the dynamics of the background magnetic texture. {If the spatial variation of the magnetic texture is sufficiently slow}, the induced current can be described by semiclassical transport theory, whereas sharp variations yield a pumping current described by the quantum Hall effect. The charge pumping effect implies that a certain amount of localized charge [Eq.~(\ref{eq:induced-charge})] is induced by the axial magnetic flux, i.e., the curl of the magnetic texture. Conversely, it also implies that a local electrostatic potential that alters the local charge distribution would induce a magnetic texture in a magnetic WSM. However, verifying the existence of such an effect remains an open question and further microscopic calculations will be required to confirm this proposal. Nevertheless, from a topological point of view, the proposed pumping current and localized charge are simple manifestations of the interplay between the real-space topology and its momentum-space counterpart, which can generally be traced back to Berry curvatures defined in the global phase space \cite{Xiao_2010}.

Secondly, by considering the coherent motion of a magnetic domain wall (DW) in a WSM, we were able to compare the pumped charge with the localized charge calculated in previous work by more direct methods \cite{Araki_DW}, and show their equivalence.
The idea of charge pumping obtained here is also applicable to all kinds of magnetic textures: 
{magnetic skyrmions and monopoles, for instance, carry pointlike charge,
whereas magnetic helices can be accompanied by arrays of localized charge, i.e., charge density waves.}
This concept may help us to design efficient spintronic devices that make use of magnetic textures in magnetic WSMs, such as in a magnetic racetrack \cite{Parkin}, where the motion of a magnetic texture can be electrically detected as a current pulse and can thus be used to read out information from an array of magnetic textures.

While we have only considered a minimal ferromagnetic toy model in this work,
the concepts developed here could be extended to other examples of TRS-broken WSMs. One particular  case of interest would be that of antiferromagnetic order in WSMs, as exhibited in $\mathrm{Mn}_3 \mathrm{Sn}$ \cite{Kubler_2016,Yang_2017,Ito}, which may exhibit similar TRS-breaking and axial vector potential effects to those presented here for ferromagnetic order. However, since the ordering is not necessarily characterized by a single order parameter, {this case would require more detailed microscopic investigations to clarify the relationship between the antiferromagnetic order and the charge degree of freedom.}

\acknowledgments{
Y.~A. is supported by JSPS KAKENHI Grant Number JP17K14316.
K.~N. is supported by JSPS KAKENHI Grant Numbers JP15H05854 and JP17K05485.
The authors would like to thank Editage (www.editage.jp) for English language editing.
}



\vspace{-12pt}
\begin{thebibliography}{99}
\vspace{-12pt}

\bibitem{Zutic_Fabian_DasSarma}
I.~\v{Z}uti\'{c}, J.~Fabian, and S.~Das Sarma,
Rev.~Mod.~Phys.~\textbf{76}, 323 (2004).

\bibitem{Brataas_Bauer_Kelly}
A.~Brataas, G.~E.~W.~Bauer, and P.~J.~Kelly,
Phys.~Rep.~\textbf{427}, 157 (2006).

\bibitem{Tatara_Kohno_Shibata}
G.~Tatara, H.~Kohno, and J.~Shibata,
Phys.~Rep.~\textbf{468}, 213 (2008). 

\bibitem{Spin_current}
S.~Maekawa, S.~O.~Valenzuela, E.~Saitoh, and T.~Kimura,
\textit{Spin Current}
(Oxford University Press, 2012).

\bibitem{Ralph}
D.~C.~Ralph and M.~D.~Stiles,
J.~Magn.~Magn.~Mater.~\textbf{320}, 1190 (2008).

\bibitem{Barnes}
S.~E.~Barnes and S.~Maekawa,
Phys.~Rev.~Lett.~\textbf{98}, 246601 (2007).

\bibitem{Young_2012}
S.~M.~Young, S.~Zaheer, J.~C.~Y.~Teo, C.~L.~Kane, E.~J.~Mele, and A.~M.~Rappe,
Phys.~Rev.~Lett.~\textbf{108}, 140405 (2012).

\bibitem{Wan_2011}
X.~Wan, A.~M.~Turner, A.~Vishwanath, and S.~Y.~Savrasov,
Phys.~Rev.~B \textbf{83}, 205101 (2011).

\bibitem{Burkov_2011}
A.~A.~Burkov and L.~Balents,
Phys.~Rev.~Lett.~\textbf{107}, 127205 (2011).

\bibitem{Zyuzin_2012}
A.~A.~Zyuzin, S.~Wu, and A.~A.~Burkov,
Phys.~Rev.~B \textbf{85}, 165110 (2012).

\bibitem{Liu_2013}
C.-X.~Liu, P.~Ye, and X.-L.~Qi,
Phys.~Rev.~B \textbf{87}, 235306 (2013).

\bibitem{Hosseini_2015}
M.~V.~Hosseini and M.~Askari,
Phys.~Rev.~B \textbf{92}, 224435 (2015).

\bibitem{Chang_2015}
H.-R.~Chang, J.~Zhou, S.-X.~Wang, W.-Y.~Shan, and D.~Xiao,
Phys.~Rev.~B \textbf{92}, 241103 (2015).

\bibitem{Araki_corr}
Y.~Araki and K.~Nomura,
Phys.~Rev.~B \textbf{93}, 094438 (2016).

\bibitem{Araki_DW}
Y.~Araki, A.~Yoshida, and K.~Nomura,
Phys.~Rev.~B \textbf{94}, 115312 (2016).

\bibitem{Grushin}
A.~G.~Grushin, J.~W.~F.~Venderbos, A.~Vishwanath, and R.~Ilan,
Phys.~Rev.~X \textbf{6}, 041046 (2016).

\bibitem{Bernevig_2016}
Z.~Wang, M.~G.~Vergniory, S.~Kushwaha, M.~Hirschberger, E.~V.~Chulkov, A.~Ernst, N.~P.~Ong, R.~J.~Cava, and B.~A.~Bernevig,
Phys.~Rev.~Lett.~\textbf{117}, 236401 (2016).

\bibitem{Hasan_2016}
G.~Chang, S.-Y.~Xu, H.~Zheng, B.~Singh, C.-H.~Hsu, I.~Belopolski, D.~S.~Sanchez, G.~Bian, N.~Alidoust, H.~Lin, and M.~Z.~Hasan,
Sci.~Rep.~\textbf{6}, 38839 (2016).

\bibitem{Kubler_2016}
J.~K\"{u}bler and C.~Felser,
Europhys.~Lett.~\textbf{114}, 47005 (2016).

\bibitem{Yang_2017}
H.~Yang, Y.~Sun, Y.~Zhang, W.-J.~Shi, S.~S.~P.~Parkin, and B.~Yan,
New J.~Phys.~\textbf{19}, 015008 (2017).

\bibitem{Ito}
N.~Ito and K.~Nomura,
J.~Phys.~Soc.~Jpn.~\textbf{86}, 063703 (2017).

\bibitem{Zyuzin_2012_2}
A.~A.~Zyuzin and A.~A.~Burkov,
Phys.~Rev.~B \textbf{86}, 115133 (2012).

\bibitem{Goswami_2013}
P.~Goswami and S.~Tewari,
Phys.~Rev.~B \textbf{88}, 245107 (2013).

\bibitem{Burkov_2014}
A.~A.~Burkov,
Phys.~Rev.~B \textbf{89}, 155104 (2014).

\bibitem{Burkov_2014_2}
A.~A.~Burkov,
Phys.~Rev.~Lett.~\textbf{113}, 187202 (2014).

\bibitem{hat}
A vector with a \textit{hat} denotes a unit vector,
i.e. $\hat{\boldsymbol{X}} = \boldsymbol{X}/|\boldsymbol{X}|$.

\bibitem{K-Y_Yang}
K.-Y.~Yang, Y.-M.~Lu, and Y.~Ran,
Phys.~Rev.~B \textbf{84}, 075129 (2011).

\bibitem{Vilenkin}
A.~Vilenkin,
Phys.~Rev.~D \textbf{22}, 3080 (1980).

\bibitem{Fukushima_2008}
K.~Fukushima, D.~E.~Kharzeev, and H.~J.~Warringa,
Phys.~Rev.~D \textbf{78}, 074033 (2008).

\bibitem{Kharzeev_2008}
D.~E.~Kharzeev, L.~D.~McLerran, and H.~J.~Warringa,
Nucl.~Phys.~A \textbf{803}, 227 (2008).

\bibitem{Kharzeev_2014}
D.~E.~Kharzeev,
Prog.~Part.~Nucl.~Phys.~\textbf{75}, 133 (2014).

\bibitem{Nandy}
S.~Nandy, G.~Sharma, A.~Taraphder, and S.~Tewari,
Phys.~Rev.~Lett.~\textbf{119}, 176804 (2017).

\bibitem{Taguchi}
K.~Taguchi and Y.~Tanaka,
Phys.~Rev.~B \textbf{91}, 054422 (2015).

\bibitem{Adler}
S.~L.~Adler,
Phys.~Rev.~\textbf{177}, 2426 (1969).

\bibitem{Bell_Jackiw}
J.~S.~Bell and R.~Jackiw,
Nuovo Cimento A \textbf{60}, 47 (1969).

\bibitem{Nielsen_Ninomiya}
H.~B.~Nielsen and M.~Ninomiya,
Phys.~Lett.~\textbf{130B}, 390 (1983).

\bibitem{Pikulin}
D.~I.~Pikulin, A.~Chen, and M.~Franz,
Phys.~Rev.~X \textbf{6}, 041021 (2016).

\bibitem{Xiao_2010}
D.~Xiao, M.-C.~Chang, and Q.~Niu,
Rev.~Mod.~Phys.~\textbf{82}, 1959 (2010).

\bibitem{Parkin}
S.~S.~P.~Parkin, M.~Hayashi, and L.~Thomas,
Science \textbf{320}, 190 (2008).

%

\end{thebibliography}
\end{document}